\documentclass[draft]{agujournal}
\usepackage{threeparttable}

\journalname{JGR-Space Physics}

\begin{document}

\title{Model of energy spectrum parameters of ground level enhancement
events in solar cycle 23}

\authors{S.-S. Wu\affil{1,2}, G. Qin\affil{3}}

 \affiliation{1}{State Key Laboratory of Space Weather, National Space
 Science Center, Chinese Academy of Sciences, Beijing 100190, China}
 \affiliation{2}{College of Earth Sciences, University of Chinese Academy
 of Sciences, Beijing 100049, China}
 \affiliation{3}{School of Science, Harbin Institute of Technology,
 Shenzhen, 518055, China}

\correspondingauthor{G. Qin}{qingang@hit.edu.cn}

\begin{keypoints}
\item Study of GLE energy spectrum parameters with conditions of the
corresponding solar events
\item Using solar conditions to determine whether there is strong
interplanetary shock acceleration
\item Preliminary prediction model of energy spectrum of ground level
enhancement events
\end{keypoints}

\begin{abstract}
Mewaldt et al. 2012 fitted the observations of the ground level enhancement
(GLE) events during solar cycle 23 to the double power-law equation to obtain
the four spectral parameters, the normalization constant $C$, low-energy
power-law slope $\gamma_1$, high-energy power-law slope $\gamma_2$, and break
energy $E_0$. There are $16$ GLEs from which we select $13$ for study by
excluding some events with complicated situation. We analyze the four
parameters with conditions of the corresponding solar events. According to
solar event conditions we divide the GLEs into two groups, one with strong
acceleration by interplanetary (IP) shocks and another one without strong
acceleration. By fitting the four parameters with solar event conditions we
obtain models of the parameters for the two groups of GLEs separately.
Therefore, we establish a model of energy spectrum of solar cycle 23 GLEs
which may be used in prediction in the future.
\end{abstract}

\section{Introduction}
\label{sec:intro}
Ground level enhancement (GLE) events are large solar energetic particle (SEP)
events in which GeV particles are able to reach the Earth's atmosphere and
produce secondary particles with intensities above the background level that
is produced by the galactic cosmic rays (GCRs) so that the counts of
ground-based neutron monitors (NMs) are enhanced \citep[e.g.,][]{Reames1999,
Lopate2006}. There are many studies on GLEs which can cause serious space
weather effects \citep[e.g.,][]{SheaEA2012}. Particularly in the solar cycle
23, there were even more GLE studies than previous periods because of more
spacecraft in orbit, including NASA's SAMPEX, ACE, STEREO and NOAA's GOES
series, onboard which the instruments extended the measurements of protons
in energy  $\sim$$0.1$ to $\sim$$500$$-$$700$ MeV \citep{MewaldtEA2012}.
In addition, SOHO and Wind spacecraft provided important observations
for GLE study in solar cycle 23, i.e., the Large Angle and Spectrometric
Coronagraph (LASCO) \citep{BruecknerEA1995} onboard SOHO to provide the
coronal images \citep[e.g.,][]{GopalswamyEA2010} and the Radio and Plasma
Wave (WAVES) experiment \citep{BougeretEA1995} onboard Wind to provide the
GLE associated radio dynamic spectra in the decameter-hectometric (DH)
wavelengths \citep[e.g.,][]{GopalswamyEA2010}. Furthermore, the mass and
kinetic energy of coronal mass ejections (CMEs) can be provided by the CME
brightness obtained from LASCO observations \citep[e.g.,][]{VourlidasEA2000,
SubramanianEA2007}, so that one can determine the strength of shock driver.
The frequency of DH type \uppercase\expandafter{\romannumeral2} bursts from
WAVES reflects the distance between interplanetary (IP) shock and the Sun
\citep[e.g.,][]{Gopalswamy2006}, so that higher starting frequency indicates
that the shock starts to accelerate particles closer to the Sun with stronger
effects. It is also indicated that lower ending frequency shows that shock
acceleration continues to larger distances from the Sun, which suggests a
strong shock. Furthermore, the composition data from the Solar Isotope
Spectrometer (SIS) \citep{StoneEA1998} onboard ACE spacecraft can be used to
determine the source of SEPs. For example, if the Fe/O ratio is higher, the
SEPs are considered from flare material instead of the shock acceleration
from solar wind or coronal materials \citep[e.g.,][]{Reames1999, CaneEA2003,
CaneEA2006}.

Some research has been done to study the GLEs detected in solar cycle 23,
of which some work focused on individual events separately
\citep[e.g.,][]{BieberEA2002, BieberEA2004, BieberEA2005, BieberEA2013,
GrechnevEA2008, McCrackenEA2008, FirozEA2012}, but others systematically
focused on the total GLEs \citep[e.g.,][]{Reames2009, GopalswamyEA2010,
GopalswamyEA2012, MewaldtEA2012}. \citet{GopalswamyEA2010} studied the flare
and CME properties, they found that the median value of flares associated
with those GLEs except GLE61 is X3.8, which is far greater than that of all
flares of solar cycle 23. For GLE61, though the flare occurred behind the
west limb and the associated flare value is not determined, there might have
a large flare because the associated CME was exceptionally fast (2465 km/s in
sky plane and $2712$ km/s deprojected). \citet{GopalswamyEA2010} also found
that the average speed of GLE associated CMEs during solar cycle 23 was as
large as $1916$ km/s in the sky plane. Note that there was no CME observation
for GLE58 when SOHO was temporarily disabled. What's more, they found that
every GLE event was accompanied by a DH type
\uppercase\expandafter{\romannumeral2} burst, which indicates strong shock
\citep{Gopalswamy2006}. \citet{GopalswamyEA2012} is the enhanced version of
\citet{GopalswamyEA2010}, and they given more information about the GLE
associated flares and CMEs in solar cycle 23. What's more, they discussed the
height of CMEs  at solar particle release (SPR) time in detail, which was
compared with the work of \citet{Reames2009}.

Based on the extended measurements of energetic protons from spacecraft,
\citet{MewaldtEA2012} tested three spectral forms for fitting proton energy
spectra of GLEs during solar cycle 23. They found that the GLE spectra are
best fitted by the double power-law shape of \citet{BandEA1993} rather than
the Bessel function shape of \citet{Ramaty1979} and the power-law with
exponential-tail shape of \citet{EllisonEA1985}. The equation of the double
power-law is given by:
\begin{linenomath*}
\begin{equation}
dJ/dE=
\begin{cases}
C\left(E/E_{\text{MeV}}\right)^{\gamma_1}\exp\left(-E/E_0\right) & \text{for $E\leq\Delta\gamma E_0$}\\
C_1\left(E/E_{\text{MeV}}\right)^{\gamma_2} & \text{for $E\geq\Delta\gamma E_0$},
\end{cases}
\label{eq:BandFun}
\end{equation}
\end{linenomath*}
where 
\begin{eqnarray}
	\Delta\gamma&=&\gamma_1-\gamma_2, \nonumber\\
	C_1&=&C\left(\Delta\gamma E_0/E_{\text{MeV}}\right)^{\Delta\gamma}\exp
	\left(-\Delta\gamma\right),\nonumber\\
    E_{\text{MeV}}&=&1\text{ MeV},\nonumber
\end{eqnarray}
where $J$ is fluence, and $E$ is kinetic energy per nucleon. It is noted that
there are four parameters in the double power-law Equation (\ref{eq:BandFun}),
namely, normalization constant $C$, low-energy power-law slope $\gamma_1$,
high-energy power-law slope $\gamma_2$, and break energy $E_0$. Since
$E_0\gg E_{\text{MeV}}$, $C$ is approximately equal to the spectrum value at
$E=E_{\text{MeV}}$. The double power-law is a piecewise function, and the
demarcation point is $\Delta\gamma E_0$, which is called transition energy
\citep[e.g.,][]{MewaldtEA2012}. If the maximum energy of measurements is less
than transition energy, the spectrum can be fitted just using the upper
expression of Equation~(\ref{eq:BandFun}), which is the Ellison-Ramaty form.
However, since there were spacecraft measurements of extended energy channels
available, it is possible for one to determine which formula to be
appropriate.

The probable causes of spectral shape are suggested as the shock
acceleration \citep[e.g.,][]{CohenEA2005, LiEA2009, TylkaEA2005, TylkaEA2006}
or transport effect \citep[e.g.,][]{LiEA2015, ZhaoEA2016}. \citet{LiEA2015}
studied the double power-law shape of GLEs in solar cycle 23 and obtained
the analytical solution of the energy spectrum from the Parker diffusion
equation with some assumptions. They found that the spectral shape of nine
western GLEs are due to the scatter-dominated transport effect, while
diffusive shock acceleration in IP space plays an important role in other
near-central meridian GLEs. On the other hand, by only considering energetic
particles' transport effects, \cite{ZhaoEA2016} adopted a simulation model
for energetic particles to study the formation of double power-law spectra.
With the spectrum of magnetic turbulence including only the inertial range,
they obtained a double power-law spectrum at 1 AU by assuming a power-law
source spectrum at the Sun, and they found that the harder the power-law
slope of turbulence is, the harder the low-energy power-law slope $\gamma_1$
is and the smaller the break energy $E_0$ is. They also suggested that
including the energy-containing range the break energy $E_0$ can be decreased,
and the smaller the correlation length in the turbulence spectrum is, the
lower the break energy $E_0$ is.

It is convenient to use the double power-law with the form in Equation
(\ref{eq:BandFun}) for the model of SEP energy spectra since there are only
four parameters needed. \citet{MewaldtEA2012} studied the fitted parameters,
i.e., $C$, $\gamma_1$, $\gamma_2$, and $E_0$, of the double power-law model
by comparing the results from the $16$ GLEs and $22$ large non-GLE events
during solar cycle 23. They found that $\gamma_1$ of GLEs is usually similar
to that of typical large SEP events, while $\gamma_2$ of GLEs is harder than
that of typical large SEP events. They also found that $E_0$ of GLEs is
usually in the range of $\sim$2 to 46 MeV. 

In this paper we analyze the spectral parameters of GLEs during solar cycle
23 with conditions of the corresponding solar events to obtain models of the
parameters, thus the energy spectra of GLEs can be provided which may be used
for prediction. In Section~\ref{sec:obser}, we introduce some observation
characteristics of the intensity-time profiles of various energy protons and
make the data selection. In Section~\ref{sec:classi}, we show the
classification of the selected GLEs. In Section~\ref{sec:analysis}, we present
the correlations of spectral parameters with each other and solar activity.
In Section~\ref{sec:model}, we obtain the statistical expressions of $C$,
$\gamma_1$, $\gamma_2$ and $E_0$, with which we establish a prediction model
of energy spectrum of GLEs. In section~\ref{sec:validity}, the modeling
results are compared with observations for six GLEs from solar cycle 22 and
one GLE from solar cycle 24. Finally, we present conclusions and discussion
in Section~\ref{sec:conclu}.

\section{Observations and Data Selection}
\label{sec:obser} 
\begin{figure}
\centering
\includegraphics[height=8in]{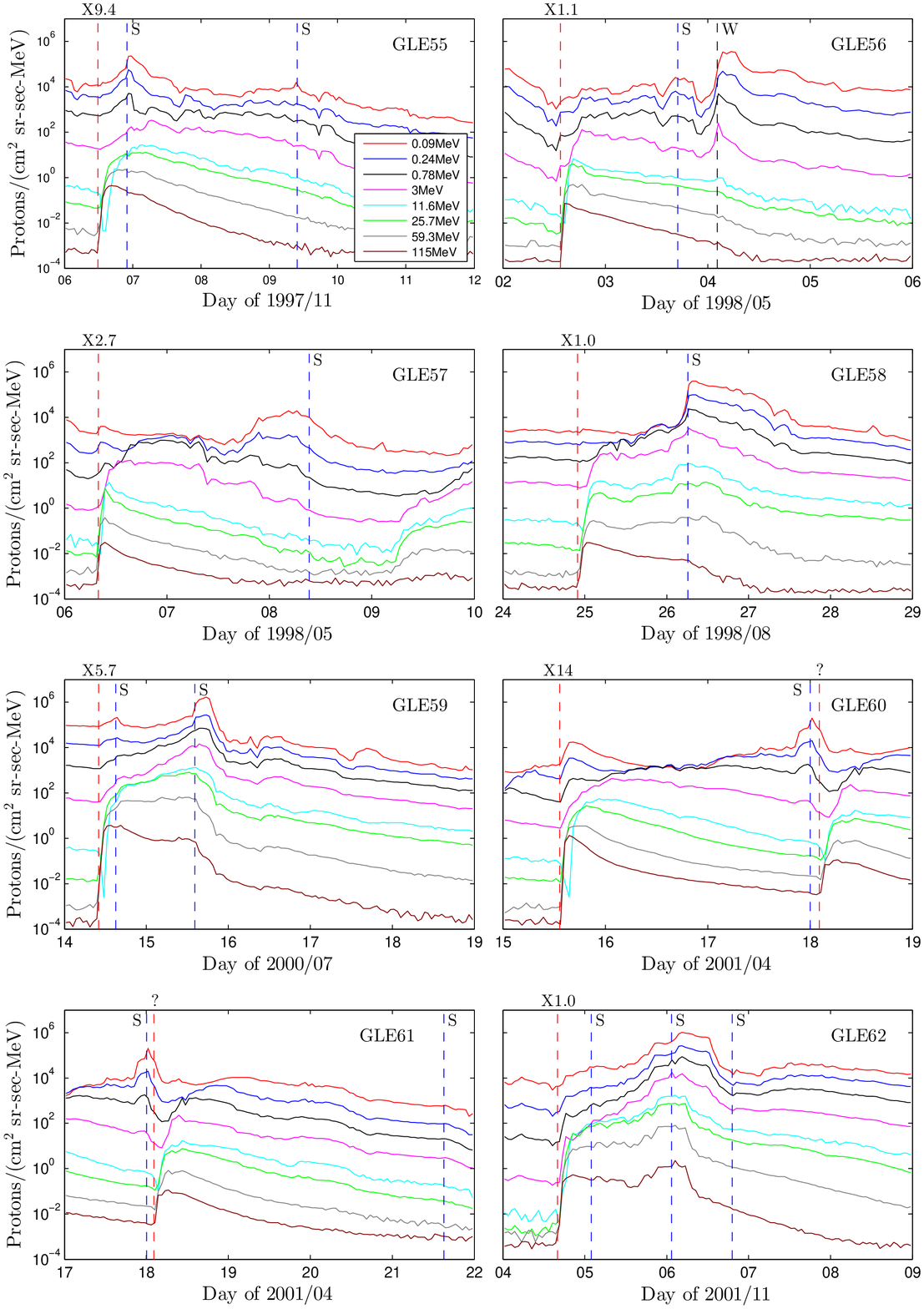}
\caption{The proton intensity-time profiles for GLE55 to GLE62. The red
and blue vertical dashed lines denote a solar flare eruption and an IP shock
arrival at 1 AU, respectively. The black vertical dashed line in the upper
right panel denotes a forward-traveling wave.}
\label{fig:timeProfile1}
\end{figure}

\begin{figure}
\centering
\includegraphics[height=8in]{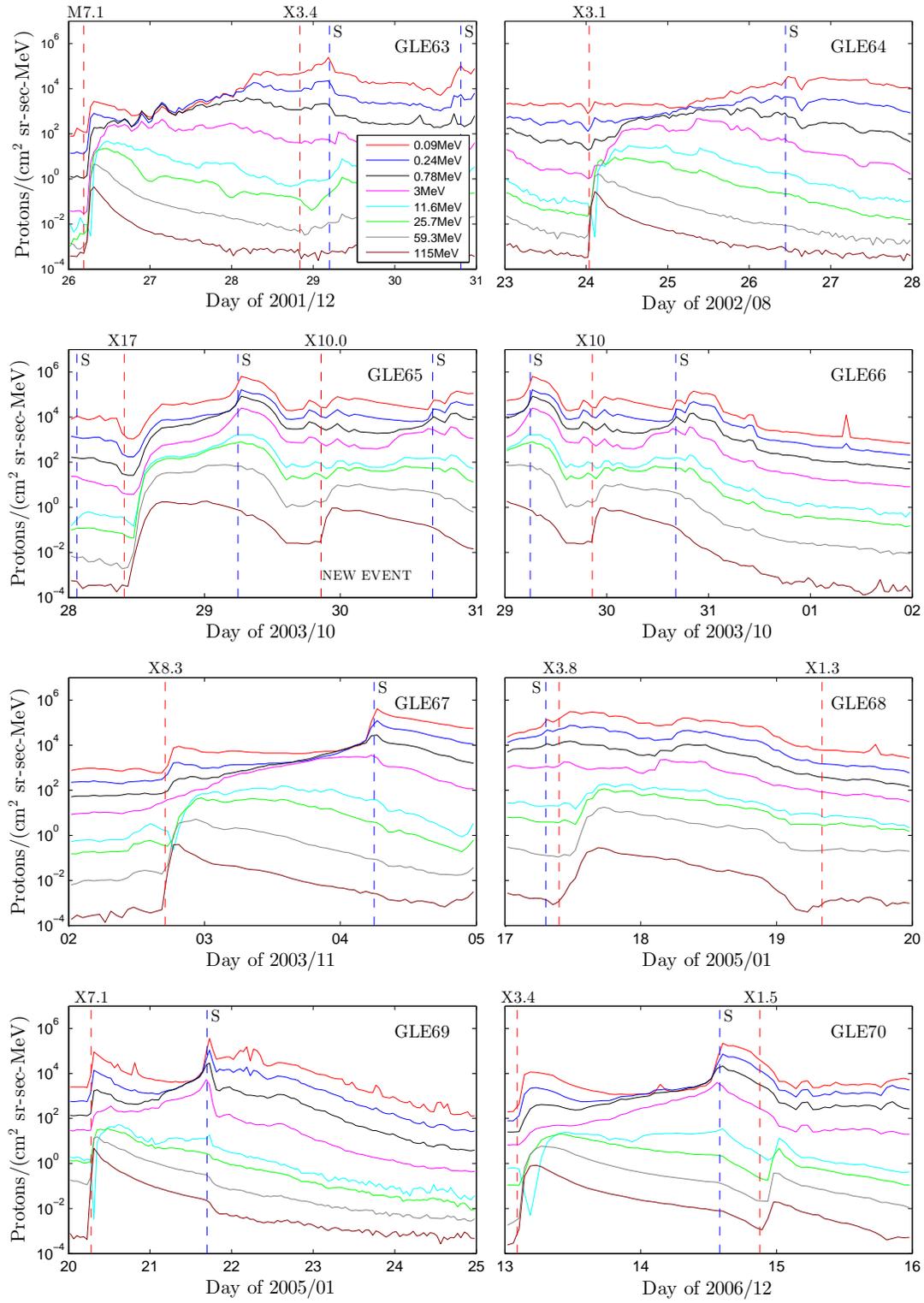}
\caption{Similar as Figure \ref{fig:timeProfile1} except for GLE63 to GLE70.
}
\label{fig:timeProfile2}
\end{figure}

In a GLE event, large amount of energetic particles are accelerated near the
solar surface or in the IP space, and they are consequently transported in the
heliosphere. The energetic particles can be measured at 1 AU by several
spacecraft simultaneously, such as ACE, GOES, SAMPEX and STEREO missions.
In Figures~\ref{fig:timeProfile1} and \ref{fig:timeProfile2}, we show the
proton intensity-time profiles for the sixteen GLEs during solar cycle 23
from the Electron, Proton, and Alpha Monitor (EPAM) \citep{GoldEA1998} on ACE
and Energetic Particle Sensor (EPS) \citep{OnsagerEA1996} on GOES-8 or
GOES-11. For GLE58, the fourth GLE in solar cycle 23 in the panel of the
second row and the second column of Figure~\ref{fig:timeProfile1}, there was
an X1.0 class flare that began at 21:50 UT on August 24, 1998 indicated by
a red vertical dashed line, and the measurements had an impulsive enhancement
after tens of minutes. When an IP shock arrived at 1 AU indicated by a blue
vertical dashed line, there was another enhancement of proton intensities,
which is called energetic storm particle (ESP) event
\citep[e.g.,][]{RaoEA1968}. The ESP event can significantly increase the
intensity of low energy protons (typically up to tens of MeVs in GLEs) with
the peak intensity higher than that caused by a flare or a corona shock.
However, ESP event usually has little influence on high energy protons, such
as tens to hundreds MeV protons. In general, the CME originated near central
meridian associates with strong ESP event. For example, with the small
longitude, GLE58 and GLE59 had strong ESP events, while GLE60 and GLE61, which
had large longitude, were hardly affected by the IP shock. The proton
intensity-time profiles at 1 AU lasts at least two days for a GLE event,
during which if an IP shock, in addition to the GLE associated one,
corresponding to an earlier solar eruption arrives at 1 AU, the proton
intensities would be influenced by the additional IP shock. For example, in
the upper left panel of Figure~\ref{fig:timeProfile1}, GLE55 associated with
an X9.4 class flare that began at 11:49 UT on November 6, 1997, and the
arrival of the first shock was at 22:02 UT on the same day, which was less than
12 hours later than the flare, thus we consider the shock is not corresponding
to the solar event of GLE55. The upper right panel of
Figure~\ref{fig:timeProfile1} exhibits the intensity-time profiles of GLE56
with an X1.1 class flare that erupted at 13:31 UT on May 2, 1998, and the
arrival of the corresponding IP shock was at 17:00 UT on 3 May. In addition to
the associated IP shock, there was a large forward-traveling wave arriving at
02:15 UT on 4 May indicated by the black vertical dashed line, which
significantly affected the low energy proton intensities, such as particles
with energy less than 10 MeV. Thus the energy spectra of GLE55 and GLE56 are
influenced, which may be the reason why their break energy $E_0$ are much
bigger than that of other GLEs. The panel of the third row and the second
column of Figure~\ref{fig:timeProfile2} shows the intensity-time profiles for
GLE68, and there was an X3.8 class flare that erupted at 09:38 UT on January
17, 2005, but without the event associated IP shock at 1 AU. However, it is
shown that about 2 hours before the solar flare associated with GLE68, an IP
shock associated with a previous solar event arrived at $1$ AU, and we can
find that the low energy proton intensity-time profiles were affected by
previous events. It can be shown that all of the GLEs except GLE68 in solar
cycle 23 had a corresponding IP shock at 1 AU. Therefore, the energy spectra
of GLE55, GLE56, and GLE68 are not usual comparing to other GLEs in solar
cycle 23, for simplicity purpose, we exclude the three GLEs for further
analysis. However, if the additional IP shock is weak, the intensity-time
profiles may not be influenced significantly. Such as GLE62 in the lower right 
panel of Figure~\ref{fig:timeProfile1}, which shows that there were three IP
shocks, and the second IP shock corresponding to GLE62 associated with a
strong ESP event, while the other two shocks had little effects on the proton
intensities. The above flare onset times and classes are from
\citet{GopalswamyEA2012}, and the shock and wave information can be found in
web (http://www-ssg.sr.unh.edu/mag/ace/ACElists/obs$\_$list.html$\#$shocks,
https://www.cfa.harvard.edu/shocks).

\begin{figure}
\centering
\includegraphics[height=3.5in]{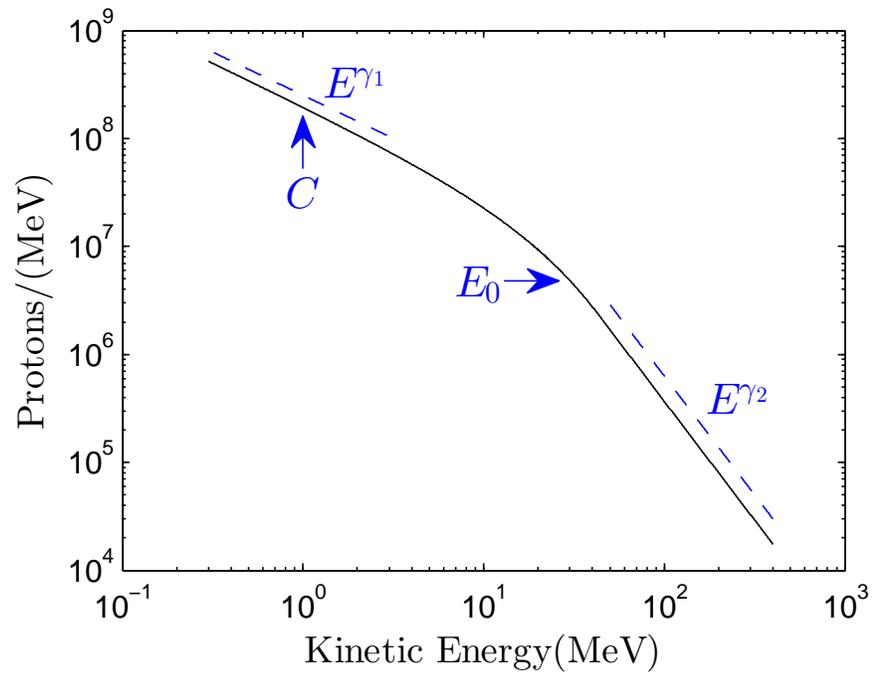}
\caption{A typical double power-law energy spectrum with the four
spectral parameters.}
\label{fig:spectrum}
\end{figure}

Energy spectrum can be obtained by integrating the proton intensity-time
profiles observed by spacecraft near the Earth. Figure~\ref{fig:spectrum} 
shows the energy spectrum of a typical GLE with the four spectral parameters.
Note that, the value of normalization constant $C$ does not affect the shape
of spectrum, but it controls the level of energy spectrum. Low-energy slope
$\gamma_1$ is always affected by the ESP event especially when the shock nose
is nearly toward the Earth, because ESP event can significantly enhance the
low energy proton intensities. It is assumed that high-energy slope $\gamma_2$
can be influenced only by a very strong toward-Earth IP shock.

In this work, the four spectral parameters of GLEs during solar cycle 23 are
obtained by \citet{MewaldtEA2012}. The flare locations listed in
Table~{\ref{tab:GLEpara}} associated to these GLEs are also from
\citet{MewaldtEA2012}. The longitude of the flares range from E09 to W117,
and the asymmetry of longitude is caused by the transport of SEPs in the
Parker spiral field. Some other related parameters that would be used in
the following sections are also listed in Table~{\ref{tab:GLEpara}}. In this
work we would analyze the four spectral parameters depending on the
event conditions for the $13$ selected GLEs in solar cycle 23. 

\section{Classification of Events}
\label{sec:classi}
\begin{figure}
\centering
\includegraphics[height=3.5in]{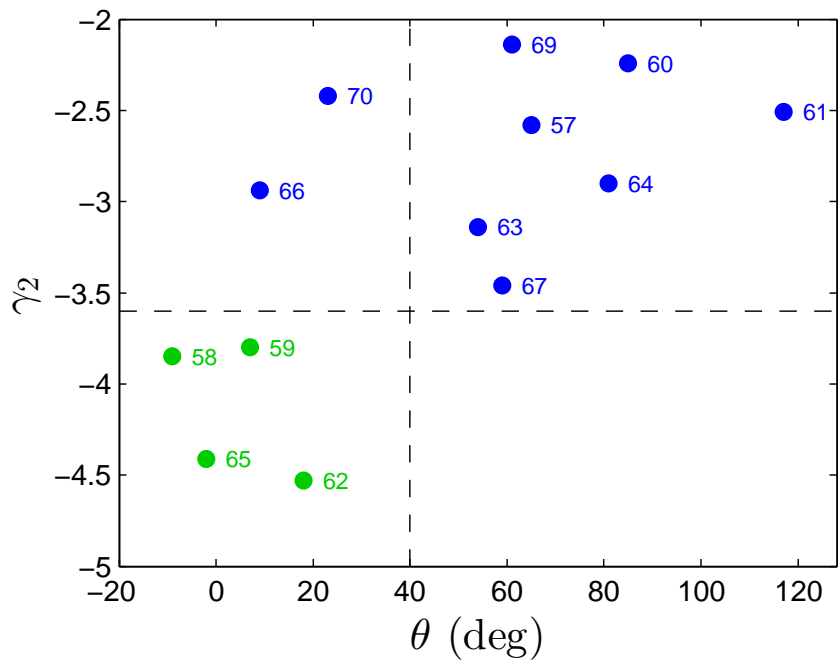}
\caption{The high-energy power-law slope $\gamma_2$ are
plotted versus flare longitude $\theta$, and the numbers in the figure
indicate the GLE number.}
\label{fig:gam2_Vs_long}
\end{figure}

Since strong IP shock acceleration of energetic particles for a GLE can
enhance the low energy part of energy spectrum significantly, the
high-energy power-law slope $\gamma_2$ would become smaller
\citep[see, e.g., Figure 13 of][]{MewaldtEA2012}.
Figure~\ref{fig:gam2_Vs_long} shows $\gamma_2$ as a function of the flare
longitude $\theta$. The numbers in the figure indicate the GLE number. We
set threshold values $\theta_t=W40$ and $\gamma_{2t}=-3.6$. The vertical
dashed line, $\theta=\theta_t$ divides all events into two parts. We assume
that $\theta\leq \theta_t$ and $\theta>\theta_t$ indicate the IP shock nose
being nearly toward the Earth and not toward the Earth, respectively. In
addition, the horizontal dashed line, $\gamma_2=\gamma_{2t}$, divides all
events into green and blue categories. From Figures~\ref{fig:timeProfile1}
and~\ref{fig:timeProfile2}, we find that for the green events the peak
intensity of $\sim$$60$ MeV protons caused by ESP event, $P_E$, is higher
than that caused by flare or coronal shock, $P_C$, or $P_E>P_C$; While
$P_E<P_C$ for the blue events. Therefore, we assume that green
($\gamma_2\leq\gamma_{2t}$) and blue ($\gamma_2>\gamma_{2t}$) events
indicate strong and weak IP shock acceleration, respectively. From
Figure~\ref{fig:gam2_Vs_long} we can see that all the events with IP
shocks being not toward the Earth have weak IP shock acceleration. However,
if the IP shock nose is toward the Earth, the events could be either in
green or blue category. We assume generally a green event corresponds to a
strong solar eruption. Therefore, it is possible we use other solar event
conditions to distinguish between the green and blue events if the shock
nose is toward the Earth.

Firstly, the brightness of CME image can be used to distinguish between the
blue and green events if the shock nose is toward the Earth. The white light
image of CME can be obtained from the LASCO onboard SOHO. The main steps are
as follows. First, we select a sequence of images which recorded the evolution
of a CME and a pre-event image from the LASCO/C2 field of view and convert
them to gray images, then use the median filtering algorithm
\citep[e.g.,][]{ShenEA2016} to get the so called "clean" images by removing
their noise. Next, we make differences of the clean images between the event
time ones and the pre-event one to obtain the pure CME images. Last, for each
of the pure CME images we calculate the average brightness, of which we choose
the largest value to denote as the brightness of CME image.

\begin{figure}
\centering
\includegraphics[height=4in]{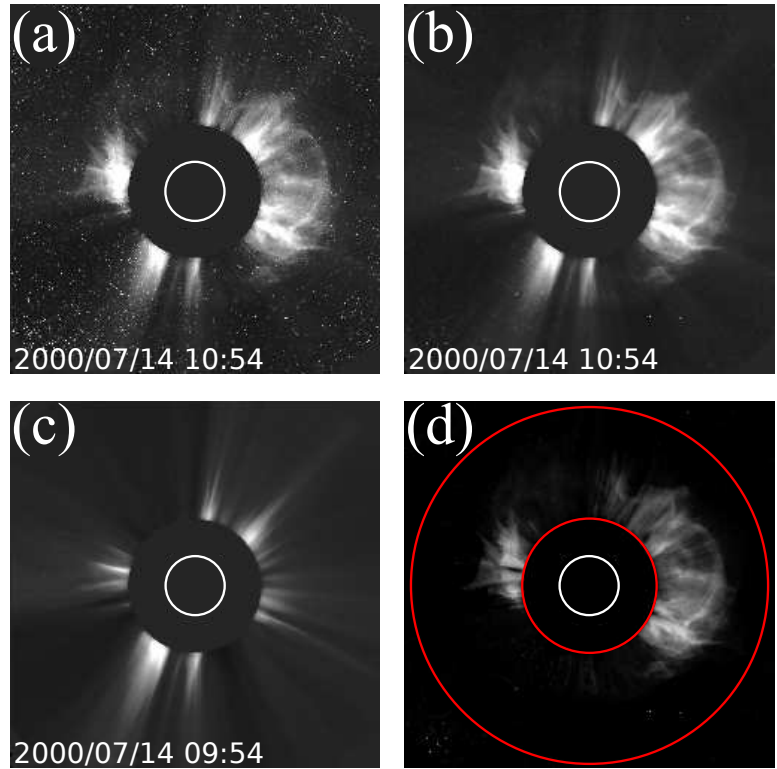}
\caption{The CME images for GLE59. Panel (a) shows the CME image that
has been converted to gray image, and the observation is from
LASCO/C2 field of view; Panel (b) is similar as panel (a) except that the
noise is removed; Panel (c) shows the pre-event image with noise removed;
Panel (d) shows the pure CME by making a difference between panel (b) and
panel (c).}
\label{fig:C2_59}
\end{figure}

\begin{figure}
\centering
\includegraphics[height=4in]{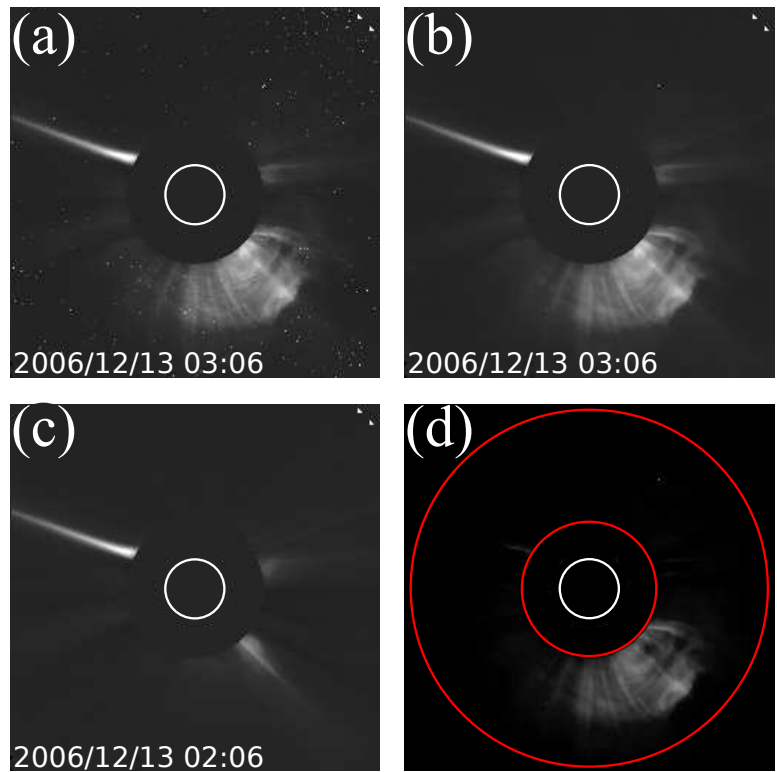}
\caption{The same as Figure \ref{fig:C2_59} except that the event is GLE70.}
\label{fig:C2_70}
\end{figure}

\begin{figure}
\centering
\includegraphics[height=3.5in]{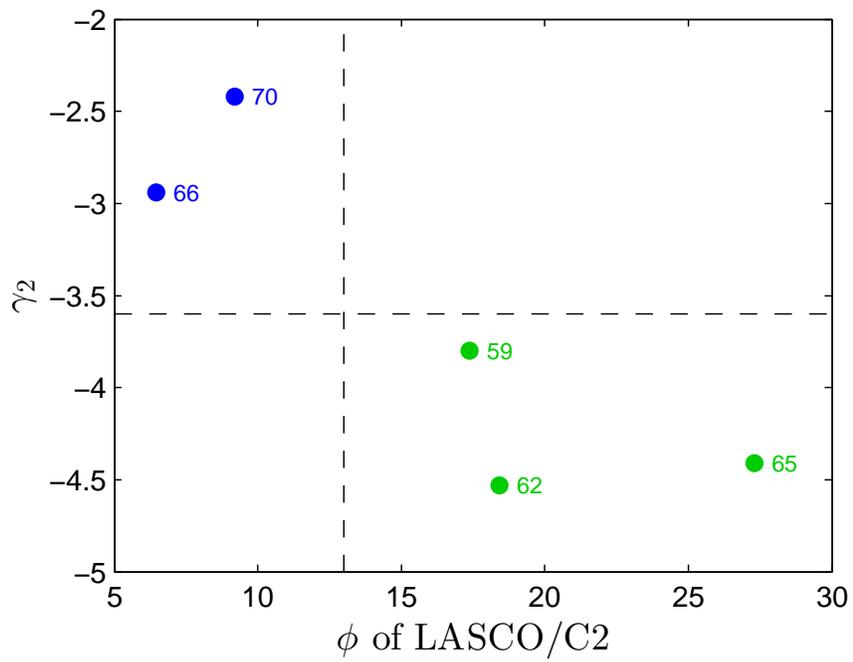}
\caption{The high-energy power-law slope $\gamma_2$ as a function of
the brightness $\phi$ of CME image of the blue and green events with
$\theta\le\theta_t$.}
\label{fig:gam2VsPhi_C2}
\end{figure}

Figure~\ref{fig:C2_59}(a) exhibits one of the gray scale CME images of GLE59,
and Figure~\ref{fig:C2_59}(b) shows the clean image by removing the noise in
Figure~\ref{fig:C2_59}(a) while Figure~\ref{fig:C2_59}(c) is the clean image
at pre-event time, and Figure~\ref{fig:C2_59}(d), which is called the pure
CME image of GLE59, is obtained by subtracting Figure~\ref{fig:C2_59}(c) from
Figure~\ref{fig:C2_59}(b). The white circle in the figure represents the size
of the Sun, and the small red circle represents the size of the occulting disk 
of C2. The average brightness is calculated in the area between the two red
circles. Figure \ref{fig:C2_70} is similar as Figure~\ref{fig:C2_59} except
that it is for GLE70. Comparing the pure CME images of GLE59 and GLE70 we find 
that the brightness of CME image of GLE70 is significantly lower than that of
GLE59. The brightness of CME image of the events with shock nose toward the
Earth are plotted as function of $\gamma_2$ in Figure~\ref{fig:gam2VsPhi_C2},
in which we denote $\phi$ as the brightness of CME image, and it's shown that
$\phi$ of green events are larger than that of blue events. Note that GLE58,
a green event, is not included in Figure~\ref{fig:gam2VsPhi_C2} since there
was no C2 data available for GLE58. Furthermore, we can set the threshold
value $\phi_{t}=13$. When the flare longitude $\theta$ of a GLE is less than
the threshold $\theta_t$, $\gamma_2$ would be smaller and larger than
$\gamma_{2t}$ indicating the strong and weak IP shock acceleration if $\phi$
is larger and smaller than $\phi_{t}$, respectively. LASCO/C3 field of view
can also record the evolution of CMEs, so that one can use C3 images to
determine the IP shock acceleration strength similarly.

\begin{figure}
\centering
\includegraphics[height=3.5in]{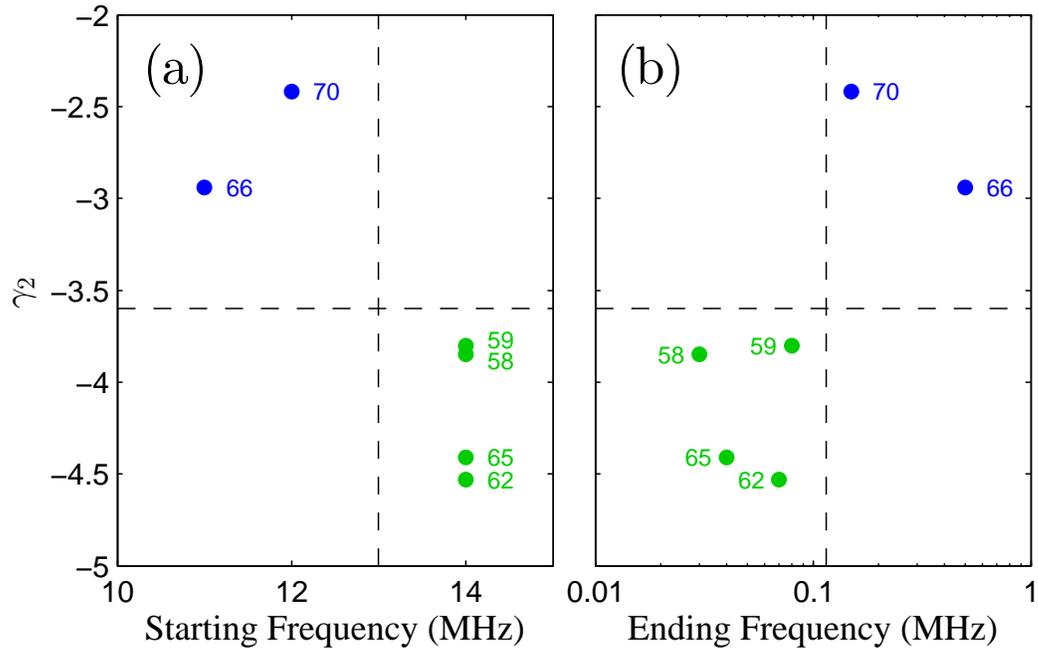}
\caption{The high-energy power-law slope $\gamma_2$ as a function of
the starting and ending frequencies of DH type
\uppercase\expandafter{\romannumeral2} radio bursts in left and right panels,
respectively, for blue and green events with $\theta\le\theta_t$. Note that
in panel (a) the starting frequencies of the green events are greater than or
equal to $14$ MHz.}
\label{fig:gam2VsType2}
\end{figure}

Next, we show that the starting and ending frequencies of DH type
\uppercase\expandafter{\romannumeral2} bursts from the WAVES onboard Wind
can also be used to separate the green and blue events if the shock nose is
toward the Earth. In Figure \ref{fig:gam2VsType2}, we show $\gamma_2$ as a
function of the starting and ending frequencies of the associated DH type
\uppercase\expandafter{\romannumeral2} bursts, obtained by
\citet{GopalswamyEA2010}, in left and right panels, respectively. It is noted
that the maximum frequency of measurements is $14$ MHz, so that the starting
frequency of the green events are all greater than or equal to $14$ MHz. We
find that both the starting and ending frequencies of DH type
\uppercase\expandafter{\romannumeral2} bursts are effective in distinguishing
the blue and green events. The frequency of DH type
\uppercase\expandafter{\romannumeral2} bursts reflects the distance between
IP shock and the Sun \citep[e.g.,][]{Gopalswamy2006}, so that higher starting
frequency indicates that the shock starts to accelerate particles closer to
the Sun. Therefore, the green events with a shorter distance from the Sun to
start accelerating may have stronger acceleration of the shock. The ending
frequencies of the green events are less than that of the blue ones, it is
suggested that the shock acceleration of green events continues to larger
distances from the Sun. However, recognizing a DH type II burst from a dynamic
spectrum might take some time, because the frequency drift rates are not very
large and the bursts are often intermittent. Therefore, the starting and
ending frequencies of DH type \uppercase\expandafter{\romannumeral2} bursts
is not suitable for predicting energy spectra.

\begin{figure}
\centering
\includegraphics[height=3.5in]{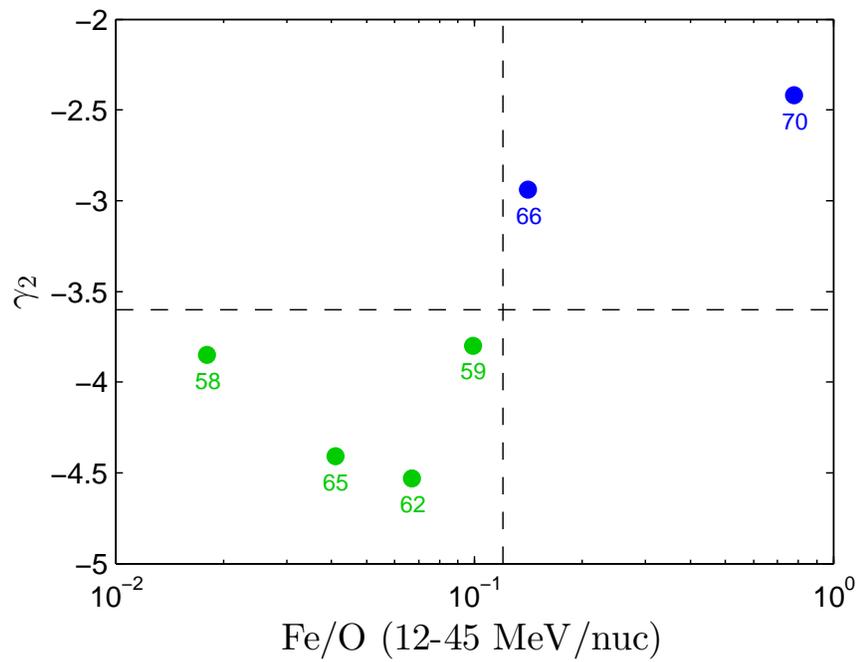}
\caption{The high-energy power-law slope $\gamma_2$ as a function of the
12$-$45 MeV/nuc Fe/O ratio measured by ACE/SIS for blue and green events
with $\theta\le\theta_t$.}
\label{fig:gam2_FeO}
\end{figure}

Furthermore, the event-integrated Fe/O ratio and Ne/O ratio can be used to
determine the acceleration source \citep[e.g.,][]{Reames1999}.
Figure~\ref{fig:gam2_FeO} shows the 12$-$45 MeV/nuc Fe/O ratio, obtained by
\citet{MewaldtEA2012}, from ACE/SIS measurements. It is shown that for
$\theta\le\theta_t$, if the 12$-$45 MeV/nuc Fe/O ratio is smaller (larger)
than $0.15$, $\gamma_2$ would be smaller (larger) than $\gamma_{2t}$,
indicating strong (weak) IP shock acceleration. It is assumed that the high
value of Fe/O ratio indicates the SEPs from flare material, while the low
value indicates the SEPs from solar wind or coronal materials
\citep[e.g.,][]{Reames1999, CaneEA2003, CaneEA2006}. The SEPs from solar
wind or coronal materials indicates stronger IP shock acceleration than
that from flare material does. We can get similar results with 12$-$45
MeV/nuc Ne/O ratio. From the above analysis, it is shown that the 12$-$45
MeV/nuc Fe/O ratio and Ne/O ratio can be used to determine if the IP shock
acceleration is strong or weak, the results are consistent with our previous
findings. In addition, other properties such as ionization states and isotope
abundances studied by \citet{Reames1999} may also distinguish these events
effectively. However, the composition data of an SEP event are too late for
one to use for predicting energy spectra.
 
Therefore, one can use the brightness of CME image to distinguish between
green and blue events if the shock nose is toward the Earth.

\section{Statistical Analysis}
\label{sec:analysis}
In order to obtain a model with good results comparing to the observations,
we first try to eliminate the correlations between the spectral parameters
themselves. Besides, one can also obtain physical understanding of the GLE
phenomenon in this way. The cross correlations of the four spectral
parameters for the thirteen selected GLEs are presented in
Figure~\ref{fig:crossCorre}. Here, blue and green stand for blue and green
events, respectively, and dashed lines indicate the linear fitting. The
regression parameters, correlation coefficients, and the level of statistical
significance of the fitting are presented in Table~\ref{tab:statiPara}.
From the cross correlations we find that only $\gamma_1$ and $\log{E_0}$
have a good correlation. Therefore, we can eliminate one parameter from
$\gamma_1$ and $E_0$ when we establish a model of energy spectrum of GLEs.

GLEs are caused by solar eruptions, the strength of which is relevant to the
solar activity. Thus, we assume that the energy spectrum is relevant to the 
solar activity. The 10.7 cm solar radio flux ($F_{10.7}$) is one of the 
indices of solar activity \citep[e.g.,][]{Tapping2013}, so that $F_{10.7}$
can be used to study the energy spectrum. Figure~\ref{fig:E0andGam2_F10}
shows the spectral parameters, $\log E_0$ and $\gamma_2$ for the selected
GLEs in the upper and lower panels, respectively, as function of $F_{10.7}$
which is the value for previous day from the NOAA's Space Weather Prediction
Center (SWPC, ftp://ftp.swpc.noaa.gov/pub/warehouse). In the upper panel of
Figure~\ref{fig:E0andGam2_F10}, $E_0$ increases with the increase of
$F_{10.7}$ for both types of events, which is reasonable since energetic
particles may be accelerated more efficiently if the solar activity is high.
In the lower panel of Figure~\ref{fig:E0andGam2_F10}, the green and
blue dashed fitting lines have the same trend, while the green and blue lines
are separated with the green line is lower in the value of $\gamma_2$ since
the strong IP shock acceleration can significantly increase the intensity of
low energy particles to make high-energy slope $\gamma_2$ smaller.

Another spectral parameter, the normalization constant $C$, is not related
to the shape of energy spectrum, but it controls the level of the energy
spectrum, i.e., $C$ is associated with the energy spectra in $1$ MeV. It is
known that the low energy protons can be influenced by  the IP shock, whose
strength is relevant with the flare longitude $\theta$. Therefore, $C$ is
possibly relevant with $\theta$.  The strength of soft X-ray burst is
relevant with solar activity at the time of event \citep{ThomasEA1971}, and
flare may also be a source of particle acceleration. Therefore, $C$ is
assumed to be also related to $F_{sxr}$, which is the integral soft X-ray
flux of flare and its value can be found in the NOAA's SWPC
(ftp://ftp.swpc.noaa.gov/pub/warehouse). Note that there is no data for
$F_{sxr}$ of GLE61 which is a backside event. One may use the space speed
of CME to determine the value of $F_{sxr}$. Figure~\ref{fig:CME_Flare} shows
the relationship between $F_{sxr}$ and CME space speed $v_{cme}$ which is
from \citet{GopalswamyEA2012}. It is clear to see that $\log F_{sxr}$ can be
expressed as a linear function of $v_{cme}$,
\[\log F_{sxr}=3.82\times 10^{-4}\frac{v_{cme}}{v_0}-1.11,\]
where $v_0=1$ km/s and the expression is used to calculate $F_{sxr}$ of GLE61.
Figure~\ref{fig:para_C_1} shows that $\log{C}$ is plotted versus $\theta$ and
$\log{F_{sxr}}$, from which we can see that both the blue and green events
have good correlations between $\log{C}$ and $\theta$ while only blue events
have a moderate correlation between $\log{C}$ and $\log{F_{sxr}}$. Therefore,
blue events may have a better correlation between $\log{C}$ and the
combination of $\theta$ and $\log{F_{sxr}}$, such as
$\theta\cdot\log{(F_{sxr}/5)}$. Choosing $\log{(F_{sxr}/5)}$ instead of
$\log{F_{sxr}}$ can make all of the values have the same sign, so that it is
easy to combine with $\theta$ for linear fit. Figure~\ref{fig:para_C_2}
exhibits the linear fitting for $\log C$ as a function of $\theta$ for green
events and that of $\theta\log\left(F_{sxr}/5\right)$ for blue events in left
and right panels, respectively. It is shown for both green and blue events
good linear fitting can be obtained.

The values of regression parameters $a$ and $b$, the correlation coefficients
and the level of statistical significance are listed in
Table~\ref{tab:statiPara}.

\section{Energy Spectrum Model for GLEs}
\label{sec:model}
From the results in Figures~\ref{fig:crossCorre}, \ref{fig:E0andGam2_F10},
\ref{fig:para_C_2}, and Table~\ref{tab:statiPara}, the expressions of the
spectral parameters are given as following:
\begin{linenomath*}
\begin{equation}
E_0 = 10^{aF_{10.7}+b},\label{eq_E0}
\end{equation}
\end{linenomath*}
with $a=0.00404$ and $b=0.360$ for blue events, and $a=0.00326$ and $b=0.634$ for green events;
\begin{linenomath*}
\begin{equation}
\gamma_2 = aF_{10.7}+b,\label{eq_gamma2}
\end{equation}
\end{linenomath*}
with $a=-0.00521$ and $b=-1.78$ for blue events, and $a=-0.00327$ and $b=-3.47$ for green events;
\begin{linenomath*}
\begin{eqnarray}
\gamma_1 &=& a\log{E_0}+b,\label{eq_gamma1}
\end{eqnarray}
\end{linenomath*}
with $a=-0.519$ and $b=-0.579$ for blue events, and $a=0.655$ and $b=-2.02$ for green events;
\begin{linenomath*}
\begin{eqnarray}
C &=&
\begin{cases}
10^{0.0252\theta+9.08}& \text{for green events;}\\
10^{0.0103\theta\log{(F_{sxr}/5)}+8.86} & \text{for blue events}.\label{eq_C}
\end{cases}
\end{eqnarray}
\end{linenomath*}
The Equations (\ref{eq:BandFun}) and (\ref{eq_E0})$-$(\ref{eq_C}) may be used
to establish a model of energy spectrum with the flare's longitude $\theta$,
integral soft X-ray flux $F_{sxr}$,  and 10.7 cm solar radio flux $F_{10.7}$
as the input. In addition, to determine the type of GLEs, the brightness of
CME image of the events is also needed.

The comparison of the energy spectra between our new model and the spacecraft
observations for the 13 selected GLEs during solar cycle 24 is presented in
Figure~\ref{fig:Pre_Spectra}. Here, $\delta$ indicates the disagreement
between the model and the observations defined as the following:
\begin{linenomath*}
\begin{equation}
	\delta=\sqrt{\frac{1}{N}\sum_{i=1}^N\left[\log{F(E_i)}-\log{f(E_i)}\right]^2},
\label{eq:error}
\end{equation}
\end{linenomath*}
where $F(E)$ and $f(E)$ are energy spectra from the model and observations,
respectively. From Figure~\ref{fig:Pre_Spectra} we can see that GLE57 has
the biggest $\delta$, and the disagreement is mainly due to the fact that
$C$ from the model is not accurate relative to the observation. On the other
hand, for GLE66 and GLE67, the model results of $C$ is not accurate but the
values of $\delta$ are smaller than that for GLE57, because in addition to the
not accurate $C$ the model also provides worse parameters $\gamma_1$ for GLE66
and $\gamma_2$ for GLE67, and the combination of some not accurate parameters
might offset each other. It is also shown that except $\gamma_2$, other
parameters from the model for GLE69 are more accurate comparing to that for
GLE67, though the values $\delta$ for GLE67 and GLE69 are similar.

\section{Partial Validity Check of Energy Spectrum Model}
\label{sec:validity}
Next, to partially check its validity, we use the new model to provide
the energy spectrum of other GLEs. Table~\ref{tab:GLEpara_2} shows the key
parameters of six GLEs in solar cycle 22, GLE40, GLE41, GLE45, GLE48, GLE52,
and GLE53, and one GLE in solar cycle 24, GLE 71, as selected for this
purpose. In Table~\ref{tab:GLEpara_2}, the first four columns are the GLE No.,
solar cycle No., GLE date, and flare location, respectively, which are from
\citet{LeEA2013} and \citep{GopalswamyEA2013} for GLEs during solar cycles 22
and 24, respectively. The last two columns are for $F_{10.7}$ and $F_{sxr}$.

In solar cycle 22, there are fifteen GLEs (GLE numbers from 40 to 54),
among which four GLEs are backside events so that $F_{sxr}$ can not be
obtained. What's more, there are five events with flare longitude less than
W40 whose blue or green categories can not be determined because the data
used in the classification methods introduced in Section~\ref{sec:classi} are
not available. The remaining events as listed in Table~\ref{tab:GLEpara_2},
are selected, of which the modeling spectra are calculated in
Figure~\ref{fig:Pre_Spectra_1}. The observations are from GOES-7 differential
channels (channels P2$-$P6, energy ranging from $6.6$ to $114$ MeV, calibrated 
by \citet{SandbergEA2014}). Generally, the modeling spectra agree well with
the observations of the GLEs, except for the GLE40, for which the observations
of intensity of low energy channels are unusually low and do not agree with
the model. Figure~\ref{fig:timeProfile_GLE40} shows the intensity-time
profiles for GLE40. It is shown that the intensities of the lower energy
P2$-$P4 channels are suppressed in the first ten hours of the events, are
assumed not accurate and marked in red in the upper left panel of
Figure~\ref{fig:Pre_Spectra_1}. The inaccuracy can be assumed to be caused
by some local structure effects, e.g., magnetospheric effects, which are
not very strong for protons $>10$ MeV normally.

Figure~\ref{fig:Pre_Spectra_2} shows, for GLE71 of solar cycle 24, the
comparison between the modeling results and the observations from ACE/EPAM
(energy ranging from $\sim$$0.1$ to $\sim$$3$ MeV) and GOES-13 differential
channels (channels P6$-$P7, energy ranging from $113.3$ to $178.5$ MeV,
calibrated by \citet{Bruno2017}), and integral channels ($>5$, $>10$, $>30$,
$>50$, $>60$, and $>100$ MeV, described in \citet{MewaldtEA2005}). It is
shown that generally the modeling results agree well with observations for
GLE71, except that in lower energy range the intensity from observations,
marked in red, is relatively higher than that from modeling. In the
intensity-time profiles exhibited in Figure~\ref{fig:timeProfile_GLE71},
there are two IP shocks, first of which corresponds to an earlier event. In 
general, IP shocks can accelerate lower energy protons to enhance their
intensity. Therefore, the intensity observations of EPAM for GLE71 are higher
than the modeling results.

All in all, the modeling results exhibited in Figures~\ref{fig:Pre_Spectra_1}
and \ref{fig:Pre_Spectra_2} show that the new model can represent the data
fits of the seven GLEs to a relatively good accuracy in energy ranging from
$\sim$$6$ to $\sim$$100$$-$$200$ MeV.

\section{Conclusions and Discussion}
\label{sec:conclu}
In this paper, we analyze the four spectral parameters, i.e., the
normalization constant $C$, low-energy power-law slope $\gamma_1$,
high-energy power-law slope $\gamma_2$, and break energy $E_0$, which are
obtained from \citet{MewaldtEA2012} by fitting the GLE observations during
solar cycle 23 to the double power-law equation. In the statistics, we exclude
GLE55, GLE56, and GLE68 out of the total of 16 GLEs in solar cycle 23 for
simplicity purpose, because the conditions of the three GLEs are complicated
comparing to that of the rest ones. We divide the selected GLEs into two types
of events according to $\gamma_2$, i.e., the blue and green events with large
and small $\gamma_2$, respectively. Because large enhancement of lower energy
particles from strong IP shock acceleration would decrease $\gamma_2$, we
assume that the small $\gamma_2$ indicates strong IP shock acceleration. We
find that all the events with large longitude are blue events with large
$\gamma_2$. But it is shown that with small longitude one need to distinguish
between blue and green events. We find that the use of the brightness of
CME image, the starting and ending frequencies of DH
type \uppercase\expandafter{\romannumeral2} bursts, and the 12$-$45 MeV/nuc
Fe/O ratio and Ne/O ratio are different between the blue and green events, so
that they can be used to distinguish the blue and green events. However, to
consider about the availability of the data during near the solar eruption
we may choose the brightness of CME image to distinguish the two types of
events with small longitude.

We find in each type of GLE events, for the spectral parameters, only
$\gamma_1$ and $E_0$ have strong linear relationship when considering the
cross correlation, and $\gamma_2$, $E_0$ are relevant with 10.7 cm solar
radio flux, $F_{10.7}$. For the green events, $C$ is correlated with flare
longitude, $\theta$. While for the blue events, $C$ has a better
relationship with the combination of $\theta$ and $F_{sxr}$, i.e.,
$\theta\log{(F_{sxr}/5)}$, where $F_{sxr}$ is the integral soft X-ray
flux of the flare. Therefore, we obtain the expressions for the four
parameters as function of solar event conditions in solar cycle 23, thus a
model of energy spectrum for GLEs in the period is established. We also
compare the energy spectra from model with the observed ones of solar cycle
23 GLEs. However, we can see that for the model there are cases with one
inaccurate parameter resulting in large error, but there are some other cases
with several inaccurate parameters without resulting in large error because
the effects of inaccuracy may offset each other. It is also noted that from
Table~\ref{tab:statiPara} we can see that three out of four computed
correlations in Equations~(\ref{eq_E0})$-$(\ref{eq_C}) for the green events
fail the test of statistical significance at the usual 5\% level, which
might be because of the too small number, $4$, of events in this category.
Finally, we obtain modeling results for six GLEs of solar cycle 22 and GLE71
of solar cycle 24 to check the validity of the model, and we find that the
model can give a relatively good results in energy ranging from $\sim$$6$ to
$\sim$$100$$-$$200$ MeV for the events with longitude greater than W$40$.
However, due to the lack of observations in solar cycle 22, the modeling
spectra of low energy part haven't been checked. In addition, because the
data needed to distinguish between green and blue categories are not
available, the modeling results for small longitude events are also not
checked.

\begin{figure}
\centering
\includegraphics[height=6in]{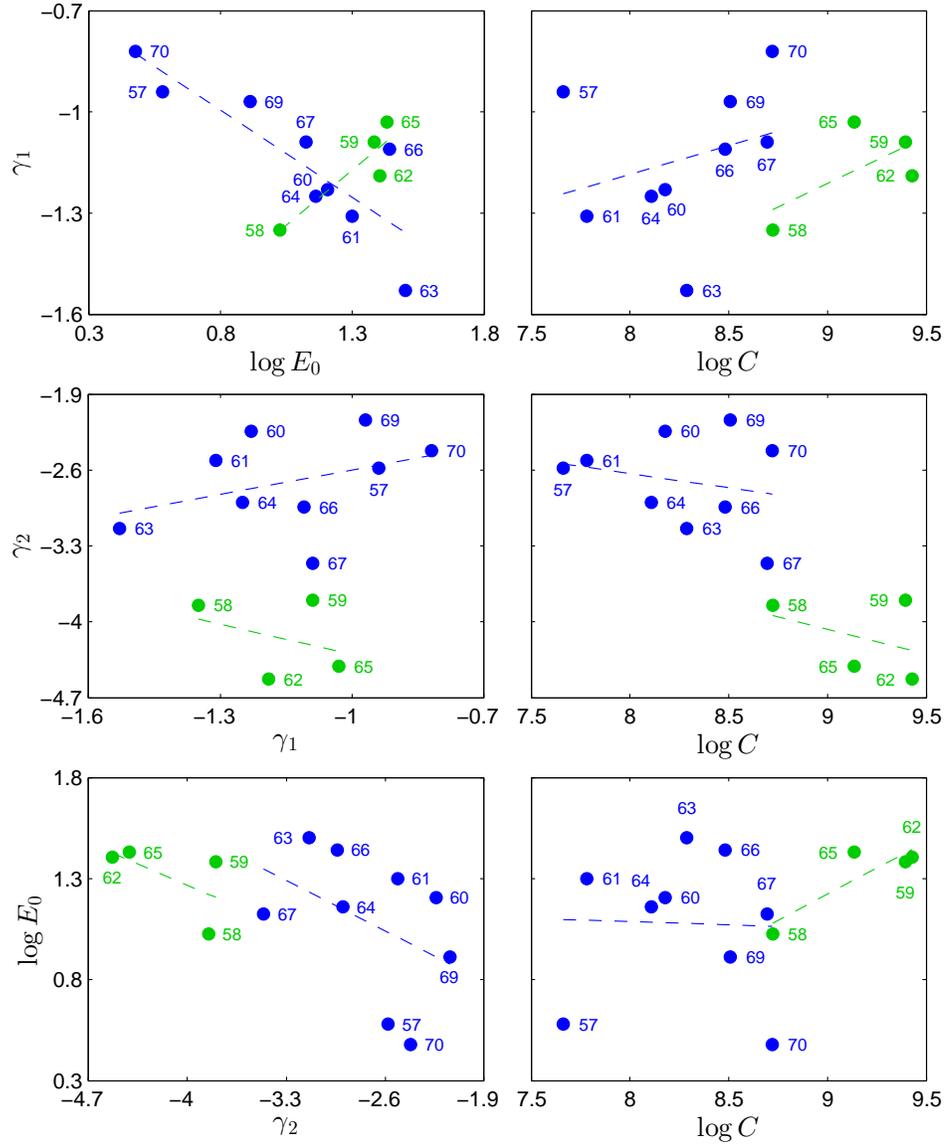}
\caption{The correlations of the four spectral parameters with each other.
Blue and green indicate the blue and green events, respectively. Dashed
lines indicate the linear fitting.}
\label{fig:crossCorre}
\end{figure}

\begin{figure}
\centering
\includegraphics[height=5in]{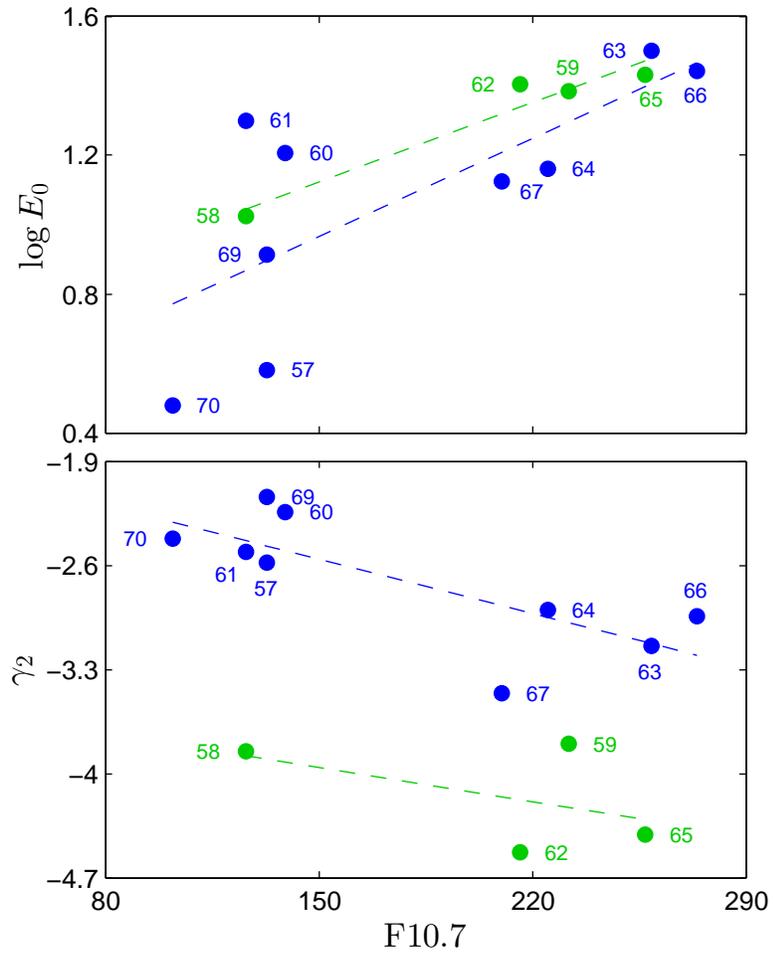}
\caption{Two spectral parameters, $\log E_0$ and $\gamma_2$ as function of the
solar activity index $F_{10.7}$ in the upper and lower panels, respectively.}
\label{fig:E0andGam2_F10}
\end{figure}

\begin{figure}
\centering
\includegraphics[height=3in]{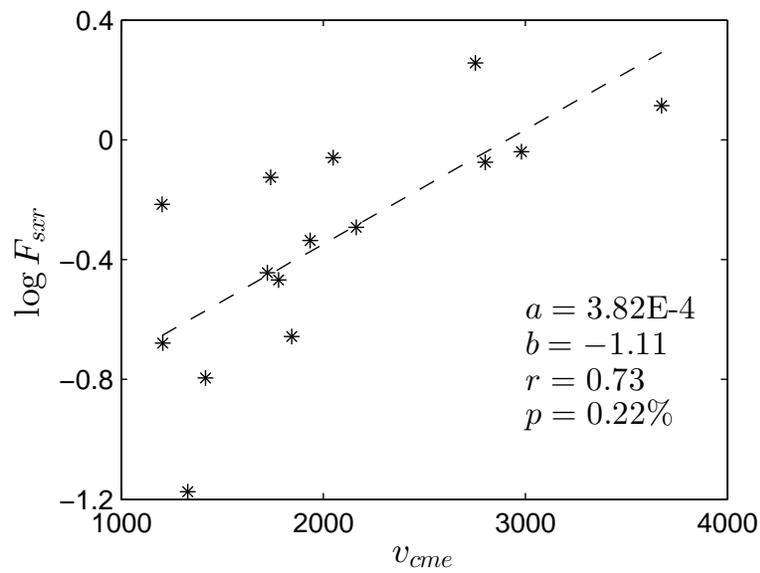}
\caption{The integral soft X-ray flux of all of the GLEs except GLE61 during
solar cycle 23, $\log{F_{sxr}}$, as a function of the space speed of CMEs,
which are from \citet{GopalswamyEA2012}. The black dashed line indicates
linear fitting. It is noted that $a$, $b$ are regression parameters and $r$,
$p$ are correlation coefficient and the level of statistical significance,
respectively.}
\label{fig:CME_Flare}
\end{figure}

\begin{figure}
\centering
\includegraphics[height=2.3in]{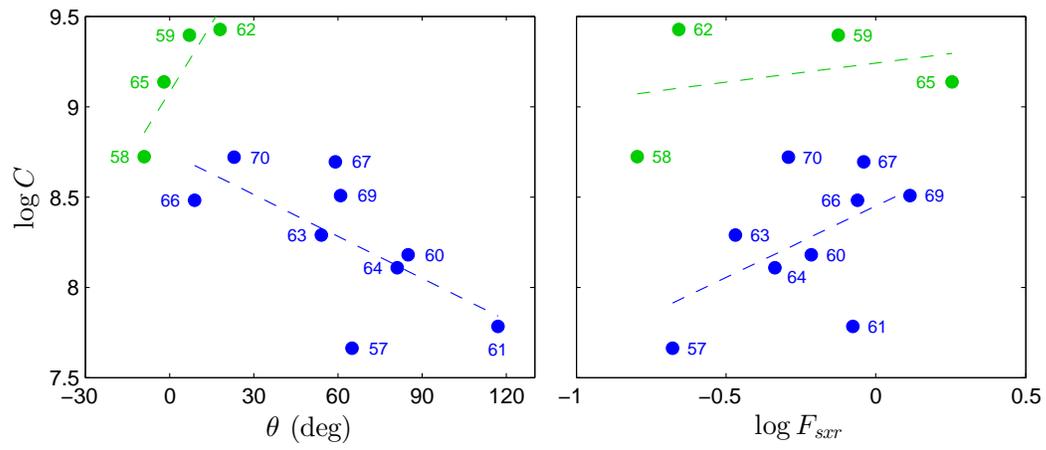}
\caption{Spectral parameter, $\log C$, is plotted versus $\theta$ and
$\log{F_{sxr}}$ in the left and right panels, respectively.}
\label{fig:para_C_1}
\end{figure}

\begin{figure}
\centering
\includegraphics[height=2.3in]{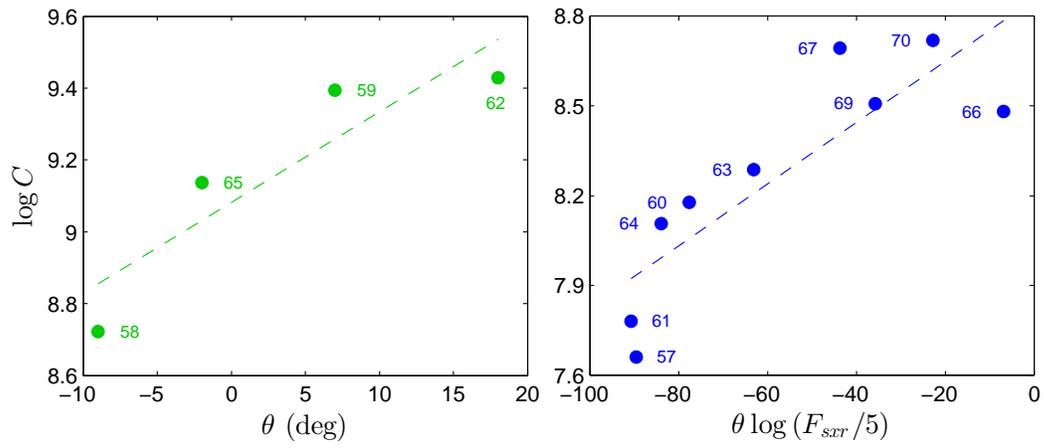}
\caption{Spectral parameter, $\log C$, is plotted as a function of $\theta$
in the left panel for green events, while $\log{C}$ of blue events is plotted
as a function of $\theta\log{(F_{sxr}/5)}$ in the right panel.}
\label{fig:para_C_2}
\end{figure}

\begin{figure}
\centering
\includegraphics[height=8in]{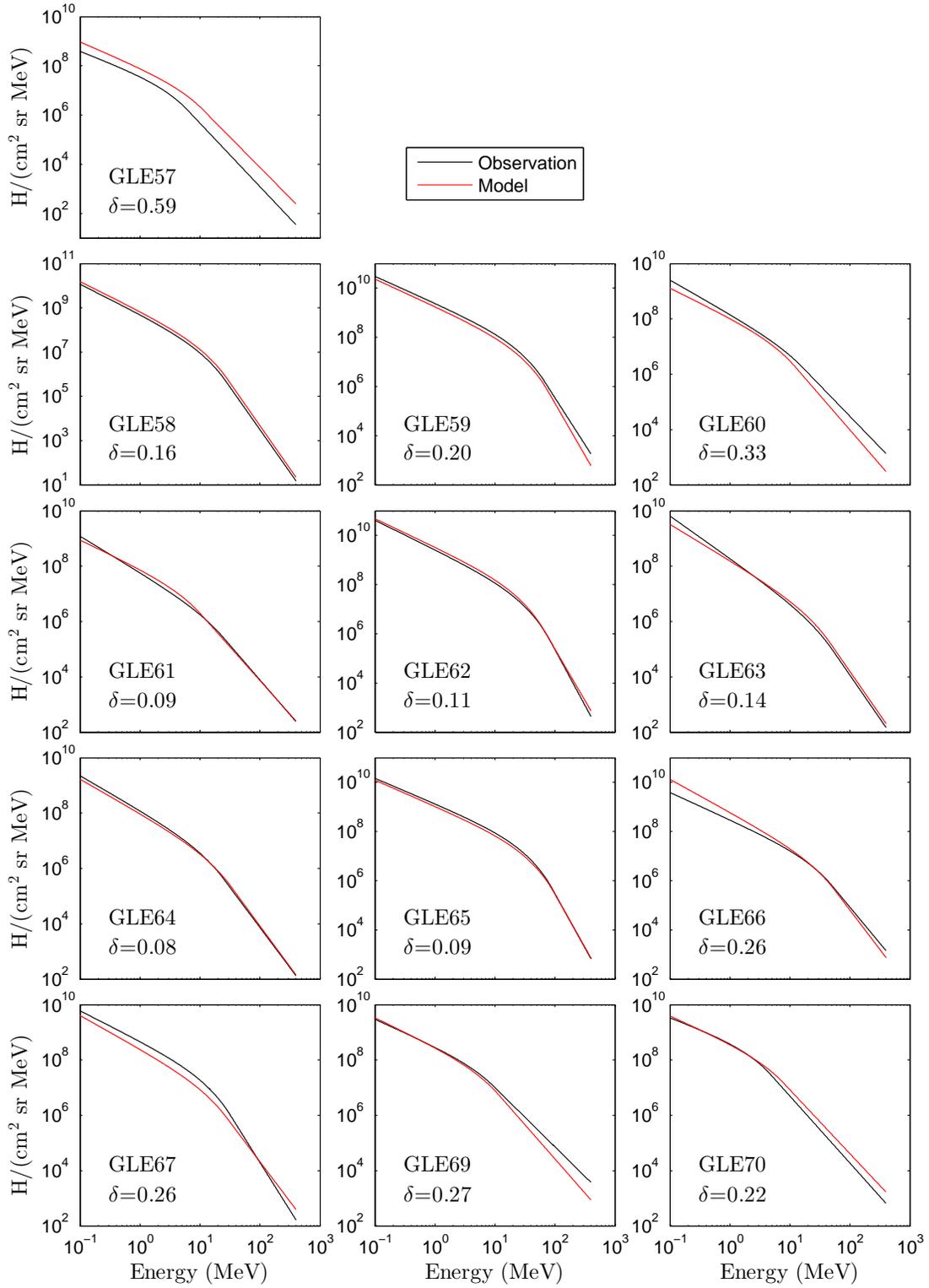}
\caption{Energy spectra from observation and model in black and red,
respectively, for 13 GLEs during solar cycle 23. $\delta$ is
calculated with equation~(\ref{eq:error}).}
\label{fig:Pre_Spectra}
\end{figure}

\begin{figure}
\centering
\includegraphics[height=3.3in]{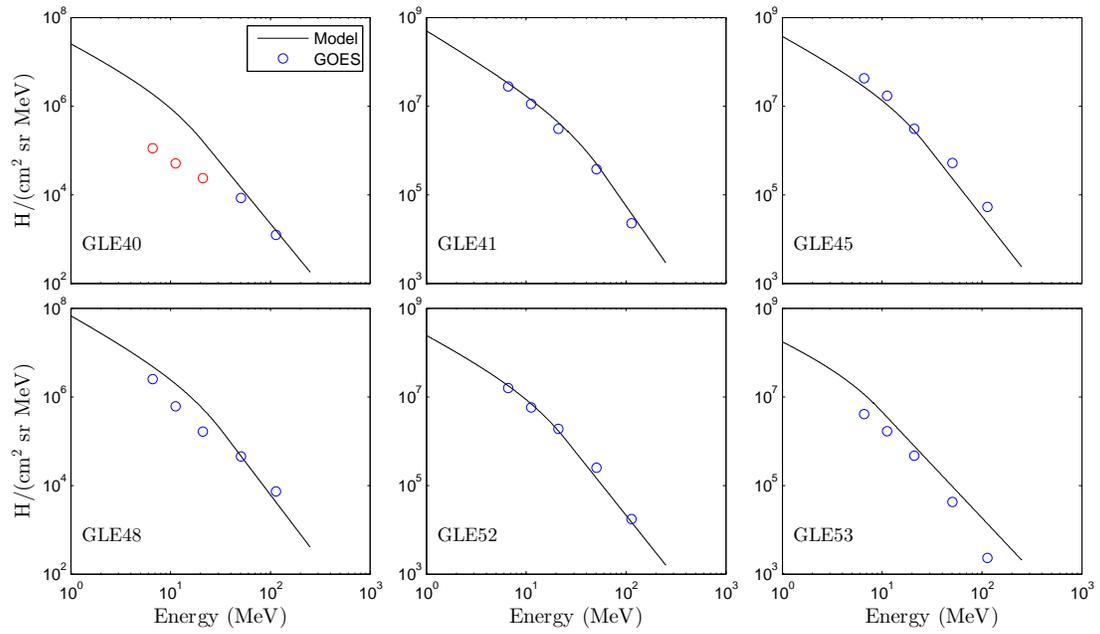}
\caption{The energy spectra from the model are compared with the observations
for six GLEs of solar cycle 22, and the observations are from GOES-7
differential channels (channels P$2$$-$P$6$). The red color indicates the
data points are influenced by some effects.}
\label{fig:Pre_Spectra_1}
\end{figure}

\begin{figure}
\centering
\includegraphics[height=3in]{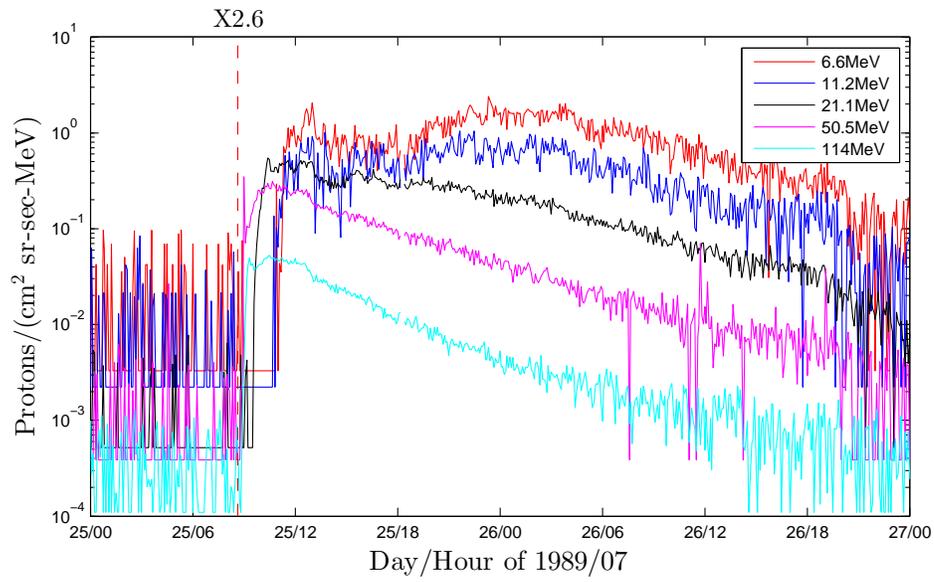}
\caption{Proton intensity-time profiles for GLE40, and the intensities of
relatively low energy protons are suppressed to some degree.}
\label{fig:timeProfile_GLE40}
\end{figure}

\begin{figure}
\centering
\includegraphics[height=2.5in]{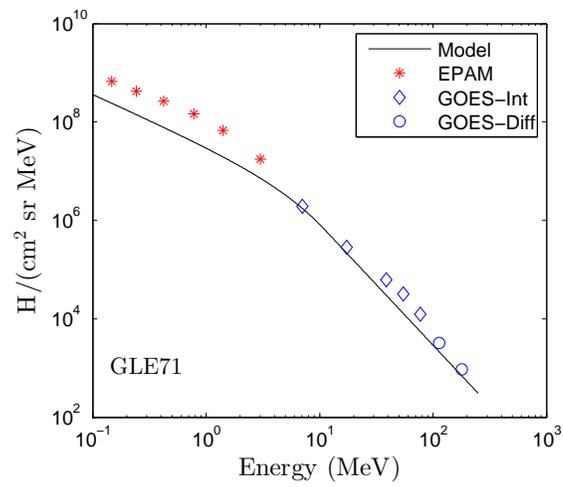}
\caption{The energy spectrum from the model are compared with the observations
for GLE71 of solar cycle 24, and the observations are from ACE/EPAM and
GOES-13 differential channels (channels P6$-$P7) and integral channels
($>5$, $>10$, $>30$, $>50$, $>60$, and $>100$ MeV). The red color indicates the
data points are influenced by some effects.}
\label{fig:Pre_Spectra_2}
\end{figure}

\begin{figure}
\centering
\includegraphics[height=3in]{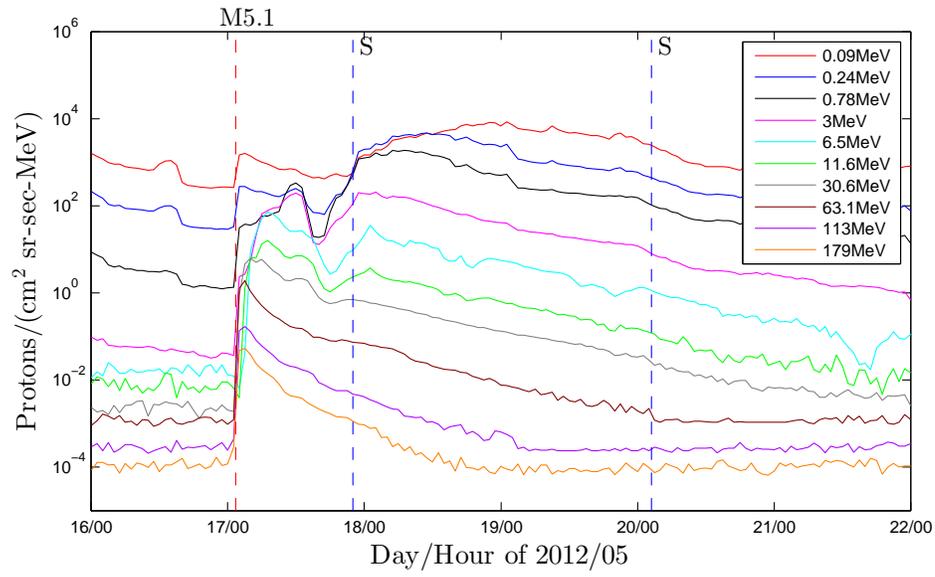}
\caption{Proton intensity-time profiles for GLE71.}
\label{fig:timeProfile_GLE71}
\end{figure}

\citet{LiEA2015} found that the spectral shape of nine western GLEs
are due to the scatter-dominated transport effect, while diffusive shock
acceleration in IP space plays an important role in other near-central
meridian GLEs. In this work, we suggest that the causes of the spectral
shape for green and blue events may be different. In fact, the slopes of
the green and blue events in the upper left panel of
Figure~\ref{fig:crossCorre} and the left panel of Figure~\ref{fig:para_C_1}
are opposite. For the upper left panel of Figure~\ref{fig:crossCorre}, we
assume the green events undergo strong IP shock acceleration to cause $E_0$
and $\gamma_1$ larger, so that $E_0$ and $\gamma_1$ are positively correlated.
For the blue events, the IP shock acceleration is weak, thus the transport
effect plays an important role in the relationship between $\gamma_1$ and
$E_0$, which is similar to the simulation results by \citet{ZhaoEA2016} who
considered only the pure transport process. As we mentioned above, $C$ is
associated with the energy spectra in 1 MeV. For the green events in the
left panel of Figure~\ref{fig:para_C_1}, the shock near Earth would
significantly increase the intensity of 1 MeV protons when the shock nose
crosses the magnetic field line connected to spacecraft, thus larger $\theta$
indicates shock nose encountering the magnetic field line earlier with higher
accelerating effects because shock strength decreases with increasing of the
solar distance. For the blue events, the situation is complex since the
fluence of the 1 MeV protons could be determined by transport effect.
Therefore, the slopes of the green and blue events may be different.

The energy spectrum model may be helpful for the future prediction of the
GLE's energy spectrum. One needs to get the inputs quickly enough after the
solar eruption. In this model the value of $F_{10.7}$ of the previous day
is used, thus the input for $F_{10.7}$ can be obtained on time. In addition,
the flare longitude $\theta$ and $F_{sxr}$ can be obtained quickly from the
solar image observed by ground-based telescopes and soft X-ray flux observed
by GOES Solar X-Ray Sensor (XRS) \citep{HanserEA1996}, respectively. If the
longitude is small, the brightness of CME image from SOHO/LASCO/C2 is needed
to determine the type of an event. However, the brightness of CME image from
SOHO/LASCO/C2 can not be obtained quickly for two reasons. Firstly, CME needs
thirteen minutes to one or two hours to transport to the view of C2. Secondly,
the data of C2 are delivered to the Earth via Deep Space Network (DSN)
stations if there is telemetry contact, but they have to wait for several
hours otherwise (see the description about the very latest SOHO images,
https://sohowww.nascom.nasa.gov/data/realtime/image-description.html).
Therefore, in order to use the energy spectrum model to predict one might have
to wait as long as several hours after the solar eruption. It is important to
study other physical parameters that can help to determine the strength of IP
shock acceleration if we want to be able to determine the type of event
quickly after the solar event.

In order to improve the model of GLE energy spectrum, among other efforts, we
need to study the transport of GLEs by comparing the spacecraft observations
with the numerical modeling \citep[e.g.,][]{QinEA2011,  QinEA2013, WangEA2012,
QiEA2017}. When we can better understand acceleration and transport effects
of energetic particles, we may be able to include the ``unusual'' GLEs
omitted in the current GLE energy spectrum model. In addition, there are
usually much more large non-GLE SEP events than GLEs in a solar cycle.
Therefore, It is also essential for us to study the energy spectra for large
non-GLE SEP events with the method similar to that in this work.

\acknowledgments
This work was supported in part by grants NNSFC 41574172, NNSFC 41374177,
and NNSFC 41125016. We thank the \textit{ACE} EPAM, SWEPAM, MAG, SIS;
\textit{GOES} EPS, XRS; \textit{SOHO} LASCO; \textit{WIND} WAVES teams for
providing the data used in this paper. The \textit{ACE} data are provided by
the ACE Science Center and the \textit{GOES} data by the NOAA. We appreciate
the availability of the \textit{WIND} data at the Coordinated Data Analysis
Web. We also acknowledge the CDAW CME catalog which is generated and
maintained at the CDAW Data Center by NASA and The Catholic University of
America in cooperation with the Naval Research Laboratory. \textit{SOHO} is
a project of international cooperation between ESA and NASA. We thank the 
anonymous referees for their extensive efforts to improve this work.

\clearpage

\clearpage

\clearpage

\clearpage

\clearpage

\clearpage

\clearpage

\clearpage

\clearpage

\clearpage

\clearpage

\clearpage

\clearpage

\clearpage

\clearpage

\clearpage

\clearpage

\clearpage

\clearpage

\clearpage
\begin{table}
\small
\caption{Some related parameters of solar cycle 23 GLEs.}
\label{tab:GLEpara}
\centering
 \begin{threeparttable}
\begin{tabular}{ccccccccccccc}
\hline
  GLE  & Date & Location & $\theta\le\theta_t$ & Category & $\phi$  &  $f_{start}$ & $f_{end}$ & Fe/O & $F_{10.7}$ & $F_{sxr}$ & $v_{cme}$\\
\# &&&&&& (MHz) & (MHz) & 12--45 MeV/n & (sfu) & (W/m$^2$) & (km/s)\\
\hline
 55\tnote{a} & 11/06/97 & S18W63 & & & & & & & 114 & 0.360\tnote{b} & 1726\\
 56\tnote{a} & 05/02/98 & S15W15 & & & & & & & 113 & 0.067 & 1332\\
 57 & 05/06/98 & S11W65 & N & blue & & & & & 133 & 0.210 & 1208\\
 58 & 08/24/98 & N35E09 & Y & green & & 14 & 0.03 & 0.018$\pm$0.001 & 126 & 0.160  & 1420\tnote{d}\\
 59 & 07/14/00 & N22W07 & Y & green & 17.39 & 14 & 0.08 & 0.099$\pm$0.010 & 232 & 0.750 & 1741\\
 60 & 04/15/01 & S20W85 & N & blue & & & & & 139 & 0.610 & 1203\\
 61 & 04/18/01 & S23W117 & N & blue & & & & & 126 & 0.838\tnote{c} & 2712\\
 62 & 11/04/01 & N06W18 & Y & green & 18.43 & 14 & 0.07 & 0.067$\pm$0.007& 216 & 0.220 & 1846\\
 63 & 12/26/01 & N08W54 & N & blue & & & & & 259 & 0.340 & 1779\\
 64 & 08/24/02 & S02W81 & N & blue & & & & & 225 & 0.460 & 1937\\
 65 & 10/28/03 & S20E02 & Y & green & 27.30 & 14 & 0.04 & 0.041$\pm$0.004 & 257 & 1.800 & 2754\\
 66 & 10/29/03 & S19W09 & Y & blue & 06.46 & 11 & 0.50 & 0.141$\pm$0.007 & 274 & 0.870 & 2049\\
 67 & 11/02/03 & S18W59 & N & blue & & & & & 210 & 0.910 & 2981\\
 68\tnote{a} & 01/17/05 & N14W25 & & & & & & & 145 & 0.840 & 2802\\
 69 & 01/20/05 & N14W61 & N & blue & & & & & 133 &1.300 & 3675\\
 70 & 12/13/06 & S06W23 & Y & blue & 09.20 & 12 & 0.15 & 0.778$\pm$0.016 & 102 & 0.510 & 2164\\
\hline
\end{tabular}
\begin{tablenotes}
        \footnotesize
\item[a] The events are excluded for statistical analysis.
\item[b] 0.036 W/m$^2$ from SWPC, but we use 0.36 W/m$^2$ according to our
calculation from GOES-9 soft X-ray flux.
\item[c] It is not actual value but derived from the correlation that is
presented in Figure~\ref{fig:CME_Flare}.
\item[d] The value is derived from ICME observations and ESA model \citep{GopalswamyEA2012}.
\end{tablenotes}
\end{threeparttable}
\end{table}

\clearpage
\begin{table}
\caption{The regression parameters ($a$, $b$), the correlation coefficients
($r$), and the level of statistical significance ($p$) of linear fittings
for the two types of events.}
\label{tab:statiPara}
\centering
	\begin{tabular}{cc|cccc|cccc}
		\hline
		\multicolumn{2}{c|}{Parameters} & \multicolumn{4}{c|}{
		Blue events} & \multicolumn{4}{c}{Green events}\\
		\cline{1-2}\cline{3-6}\cline{7-10}
		$x$ & $y$ & $a$ & $b$ & $r$ & $p$ & $a$ & $b$ & $r$ & $p$\\
		\hline
		$\log{E_0}$ & $\gamma_1$ & -0.519 & -0.579 & -0.86 & 0.30\% & 0.655 & -2.02 & 0.90 & 10\%\\
		$\log{C}$ & $\gamma_1$ & 0.171 & -2.55 & 0.30 & 44\% & 0.278 & -3.71 & 0.65 & 35\%\\
		$\gamma_1$ & $\gamma_2$ & 0.754 & -1.84 & 0.38 & 32\% & -0.934 & -5.24 & -0.35 & 65\%\\
		$\log{C}$ & $\gamma_2$ & -0.263 & -0.528 & -0.23 & 56\% & -0.456 & 0.0307 & -0.40 & 61\%\\
		$\gamma_2$ & $\log{E_0}$ & -0.357 & 0.113 & -0.44 & 24\% & -0.300 & 0.0687 & -0.59 & 41\%\\
		$\log{C}$ & $\log{E_0}$ & -0.0319 & 1.34 & -0.034 & 93\% & 0.516 & -3.42 & 0.88 & 12\%\\
 		$F_{10.7}$ & $\log{E_0}$ & 0.00404 & 0.360 & 0.73 & 2.6\% & 0.00326 & 0.634 & 0.97 & 3.0\%\\
 		$F_{10.7}$ & $\gamma_2$ & -0.00521 & -1.78 & -0.77 & 1.6\% & -0.00327 & -3.47 & -0.50 & 50\%\\
 		$\theta$ & $\log{C}$ & -0.00772 & 8.74  & -0.66 & 5.1\% & 0.0252 & 9.08 & 0.90 & 9.7\%\\
 		$\log{F_{sxr}}$ & $\log{C}$ & 0.790 & 8.45  & 0.52 & 16\% & 0.213 & 9.24 & 0.32 & 68\%\\
 		$\theta\log{\left(F_{sxr}/5\right)}$ & $\log{C}$ & 0.0103  & 8.86 & 0.85 & 0.40\%\\
		\hline
	\end{tabular}
\end{table}

\clearpage
\begin{table}
\caption{Some key parameters of solar cycles 22 and 24 GLEs.}
\label{tab:GLEpara_2}
\centering
 \begin{threeparttable}
\begin{tabular}{cccccc}
\hline
GLE &Solar & Date & Location & $F_{10.7}$ & $F_{sxr}\tnote{a}$\\
\#  &Cycle \# & & & (sfu) & (W/m$^2$)\\
\hline
40 &22& 25/07/89 & N25W84 & 186 & 0.115\\
41 &22& 16/08/89 & S18W84 & 278 & 3.451\\
45 &22& 24/10/89 & S30W57 & 211 & 1.842\\
48 &22& 24/05/90 & N33W78 & 228 & 0.282\\
52 &22& 15/06/91 & N33W69 & 201 & 1.211\\
53 &22& 25/06/92 & N09W67 & 118 & 0.801\\
71 &24& 17/05/12 & N11W76 & 131 & 0.099\\
\hline
\end{tabular}
\begin{tablenotes}
\footnotesize
\item[a] The values are obtained from GOES-7 soft X-ray flux for GLEs during solar
cycle 22. For GLE71, the value is from SWPC.
\end{tablenotes}
\end{threeparttable}
\end{table}

\listofchanges

\end{document}